# The Mass of Quasar Broad Emission Line Regions

J.A. Baldwin

*Physics and Astronomy Department, Michigan State University, 3270 Biomedical Physical Sciences Building, East Lansing, MI 48824*

G.J. Ferland

*Department of Physics and Astronomy, University of Kentucky, 177 Chemistry/Physics Building, Lexington, KY 40506*

K.T. Korista

*Department of Physics, University of Western Michigan, 1120 Everett Tower, Kalamazoo, MI 49008-5252*

F. Hamann and M. Dietrich

*Department of Astronomy, University of Florida, 211 Bryant Space Center, Gainesville, FL 32611-2055*

baldwin@pa.msu.edu

## Abstract

We show that the mass of ionized gas in the Broad Line Regions (BLRs) of luminous QSOs is at least several hundred $M_{sun}$, and probably of order $10^3$-$10^4$ $M_{sun}$. BLR mass estimates in several existing textbooks suggest lower values, but pertain to much less luminous Seyfert galaxies or include only a small fraction of the ionized/emitting volume of the BLR. The previous estimates also fail to include the large amounts of BLR gas that emit at low efficiency (in a given line), but that *must* be present based on reverberation and other studies. Very highly ionized gas, as well as partially ionized and neutral gas lying behind the ionization zones, are likely additional sources of mass within the BLR. The high masses found here imply that the chemical enrichment of the BLR cannot be controlled by mass ejection from one or a few stars. A significant stellar population in the host galaxies must be contributing. Simple scaling arguments based on normal galactic chemical enrichment and solar or higher BLR metallicities show that the *minimum* mass of the enriching stellar population is of order 10 times the BLR mass, or > $10^4$-$10^5$ $M_{sun.}$ More realistic models of the chemical and dynamical evolution in galactic nuclei suggest that much larger, bulge-size stellar populations are involved.

*Subject headings:* quasars: emission lines, galaxies: active

## 1. Introduction

Quasars are extremely luminous objects not only in their continuum radiation but also in their emission lines. Study of their emission line spectra offers the opportunity to measure, for the Broad Lined Region (BLR) gas that forms the emission lines, chemical abundances out to very high redshifts (Hamann & Ferland 1993, 1999).

Our current understanding is that quasars are powered by the release of gravitational potential energy as gas falls from the central parts of a surrounding (proto)galaxy onto an accretion disk and then into a central massive black hole (e.g. Volonteri et al. 2002; Kauffmann & Haehnelt 2000; Silk & Rees 1998). It is thus very plausible that the gas in the BLR started out as the ISM in the central region of the surrounding (proto)galaxy. The chemical abundances of the BLR gas could then tell us much about the status of chemical and stellar evolution in the inner parts of the host galaxy.

However, this claim depends on the mass of the BLRs in the high-redshift quasars that are of most interest in this type of study. If the masses are only ~1 $M_{sun}$ or less, a mass-loss episode from just one or a few stars could temporarily determine the chemical abundances in the BLR of an individual object, so that we would learn nothing about typical conditions in the central part of the host galaxy. We will show here that in fact the BLR mass in the high-redshift quasars of interest must be at least many hundreds of solar masses. Furthermore, unless the BLR is finally tuned so as to emit with maximum efficiency per unit mass, the true mass is $M_{BLR} \sim 10^3$–$10^4$ $M_{sun}$ or quite possibly even higher.

BLR mass estimates are based on the observed luminosity in the emission lines. The luminosity of a line $L_{line}$ is proportional to $n^2 V \propto n M_{BLR}$, where $n$ is a density and $V$ the emitting volume. So, for a permitted emission line, the BLR mass $M_{BLR}$ scales as ($L_{line} / n$) up to $n \sim 10^{11}$ cm$^{-3}$, above which the emission lines become thermalized (Peterson 1997; Ferland et al. 1992).

## 2. Some Textbook Examples

To give the flavor of current estimates of $M_{BLR}$, we will first turn to some of the standard textbooks about AGN and quasars. It used to be thought that, in the BLR, $n \sim 10^9$ cm$^{-3}$, based on the presence of the C III] λ1909 recombination line which has a critical density of ~3×10$^9$ cm$^{-3}$. Osterbrock (1989, p. 329) used this density and an Hβ luminosity $L_{H\beta} \sim 2 \times 10^{44}$ erg s$^{-1}$ to estimate a BLR mass $M_{BLR} \sim 2 \times 10^3$ $M_{sun}$.

However, reverberation observations of the BLR later showed that gas exists over a range in radii and that $n \sim 10^{11}$ cm$^{-3}$ is more representative of the gas forming the permitted lines (see Peterson 1993), even though some lower density gas must also be present to form C III]. Peterson (1997, p. 73) used this higher density and the C IV λ1549 luminosity $L_{42}$(C IV) (in units of 10$^{42}$ erg s$^{-1}$) to estimate $M_{BLR} \sim 10^{-3} L_{42}$(C IV) $M_{sun}$, corresponding to "even for the most luminous AGNs, less than 10 $M_{sun}$".

Krolik (1999, eqn. 10.93) gives $M_{BLR} \sim 2$ ($C$/0.1) ($N_{ionized}/3\times10^{22}$ cm$^{-2}$) $L_{ion,45}$ $\Xi^{-1} n_{H,10}^{-1}$ $M_{sun}$. Here $C$ is the covering factor $\Omega/4\pi$, $N_{ionized}$ is the column density of ionized gas, $L_{ion,45}$ is the ionizing continuum luminosity in units of 10$^{45}$ erg s$^{-1}$, $\Xi$ is an ionization parameter defined as the ratio of radiation to gas pressures, and $n_{H,10}$ is the density of hydrogen in units of 10$^{10}$ cm$^{-3}$.

A cursory reading of these texts would leave the impression that $M_{BLR}$ is anywhere from 2 to 2000 $M_\odot$. However, there are different choices of luminosity and density (and also of other parameters) buried in these mass estimates. Here we wish to refer all mass estimates to the luminosities of moderate to high-redshift QSOs found in surveys like 2DF (Boyle et al. 2000) and the Large Bright Quasar Survey (Hewett, Foltz & Chaffee 1995), since these are the objects of most interest in trying to trace the evolution of chemical abundances at large lookback times. We will take as representative of these objects a monochromatic continuum luminosity at rest wavelength 1450Å of $L_{1450} \sim 10^{44}$ erg s$^{-1}$ Å$^{-1}$, which is about a decade below the highest luminosities of high-redshift QSOs for which useful spectra are currently available, although about a half-decade above the mean of the luminosities of the redshift $z = 2$ objects available in the literature and studied by, for example, Dietrich et al (2002). To relate this continuum luminosity to the luminosities in specific emission lines of interest, we will assume that $\lambda F_\lambda$ is constant, and that the lines have equivalent widths EW(Ly$\alpha$) = 56Å, EW(C IV 1549) = 40Å and EW(H$\beta$) = 90Å (from the best fitting Baldwin Effect slopes found by Dietrich et al 2002). These will lead to an underestimate of the BLR masses of low-luminosity objects by up to a factor of 3 because they do not take into account the variation of emission line equivalent widths with luminosity (the Baldwin Effect).

In addition, to determine the connection between $L_{1450}$ and $L_{ion}$, we will use the "baseline" ionizing continuum shape adopted by Korista et al. (1997). For this continuum shape, the total number of ionizing photons is $Q(H) = 7.95 \times 10^{13} L_{1450}$ s$^{-1}$ and the total ionizing luminosity is $L_{ion} = 5650 L_{1450}$ erg s$^{-1}$, where we have taken $\lambda F_\lambda$ = constant for wavelengths larger than 1215Å. It is seen that Krolik's fiducial luminosity corresponds to $L_{1450} = 2 \times 10^{41}$ erg s$^{-1}$ Å$^{-1}$, appropriate for a Seyfert galaxy rather than for a high-redshift QSO.

When the textbook BLR mass estimates are scaled to $L_{1450} = 10^{44}$ erg s$^{-1}$ Å$^{-1}$ and $n = 10^{11}$ cm$^{-3}$ using the above relations, the new estimates become $M_{BLR}$ = 290 M$_{sun}$ (Osterbrock), 3 M$_{sun}$ (Peterson), and 100 M$_{sun}$ (Krolik, taking his other scaling factors at the fiducial values given in his equation).

However, there are a number of further improvements that should be made to these mass estimates and which will be discussed in the following sections.

## 3. Improved Mass Estimates for Single Clouds

We first estimate the minimum mass required to produce the observed emission-line luminosities from a photoionized gas cloud of constant density and illuminated by some particular ionizing flux. Peterson (1997) estimated the mass of C$^{+3}$ required to produce the observed CIV $\lambda$1549 emission line, then scaled up by the H/C abundance ratio. A better approach is to use the luminosity of Ly$\alpha$, which is formed over a much larger volume within a single cloud, and over a more extended range of cloud parameters within the BLR.

The amount of gas needed to produce the observed Ly$\alpha$ line is $M_{BLR} = ((\mu_H m_H)/(n_e \alpha_{eff-Ly\alpha}))(L_{Ly\alpha}/(h\nu_{Ly\alpha}))$, where $n_e$ is the electron density, $\mu_H = 1.42$ is the mean mass per hydrogen atom, $m_H$ is the mass of the hydrogen atom, and $\alpha_{eff-Ly\alpha}$ is the effective recombination coefficient. For the simplest case in which all Ly$\alpha$ photons are due to recombinations and all escape the BLR, $\alpha_{eff-Ly\alpha} \sim \alpha_{H-Case B}$ = 2.59–1.43 $\times 10^{-13}$ cm$^3$ s$^{-1}$ (over the electron temperature range $T_e$ = 10,000 – 20,000 K; Osterbrock 1989, Table 2.8). This yields the smallest possible mass because emission lines are assumed to be 100% efficient radiators. For $T_e$ = 20,000 K, this mass is $M_{BLR}$ = 5.1 $(10^{11}/n_e)$ $(L_{Ly\alpha}/10^{45})$ M$_{sun}$, or in terms of the continuum luminosity, $M_{BLR}$ = 29 $(10^{11}/n_e)$ $(L_{1450}/10^{44})$ M$_{sun}$.

The true Lyα emissivities tend to be lower, and so the mass larger, because of the effects of thermalization and destruction by background opacities like the Balmer continuum (Rees, Netzer & Ferland 1989). To account for this, we used the photoionization code *Cloudy* (Ferland 2002) to compute the actual Lyα emissivity from a single "fiducial" cloud with solar abundances, a hydrogen density of $n_H = 10^{11}$ cm$^{-3}$ (chosen to be the same as used for the Peterson estimate) and a flux of ionizing photons of $\Phi = 10^{19.5}$ cm$^{-2}$ s$^{-1}$ chosen to produce an ionization parameter of $U = \Phi/n_H c = 10^{-2}$, which produces the maximum equivalent width of C IV at this gas density. All other parameters were the same as those in the fiducial model of Korista et al. (1997). The calculated emissivity per unit volume as a function of depth into the cloud for a few important emission lines is shown in Figure 1. We used the actual computed volume emission at the point where Lyα was its most efficient radiator and converted these back into emissivities to find $\alpha_{eff-Ly\alpha} \sim 3.9 \times 10^{-14}$ cm$^3$ s$^{-1}$. The mass of gas required to produce Lyα is then $M_{BLR} = 100 \, (10^{11}/n_e) \, (L_{1450}/10^{44})$ M$_{sun}$.

This is considerably larger than Peterson's mass estimate from the C IV line, which in the same notation is $M_{BLR} \sim 4 \, (10^{11}/n_e) \, (L_{1450}/10^{44})$ M$_{sun}$. The reason is that Lyα is emitted over a 10-20 times larger volume than CIV (see Fig. 1). This effect can be seen in a wide variety of models of the BLR. For example, averaged over the entire cloud, the standard model of Kwan & Krolik (1981) has 1.3% of the C in the form of $C^{+3}$ as compared to 27% of the H in the form of $H^+$, while for the fiducial model used here these values are 0.8% and 12%, respectively. So, for our fiducial model, the Peterson estimate should be multiplied by 1/0.008 and our Lyα estimate by 1/0.12 to account for regions these lines do not probe. However, these corrections are sensitive to the total column density of the cloud, since extensive neutral regions exist. We will discuss this issue further in §4, below.

A better way to include the full extent of the ionized zone in this sort of model is to calculate the mass within the ionized depth of a gas cloud at a known radial distance from the central continuum source. This is the method used by Krolik (1999). *Cloudy* finds that a cloud of solar abundances with a density of $10^{11}$ cm$^{-3}$ and an ionization parameter $U = 0.033$ has an ionized column density in hydrogen $N_{ionized} = 1.7 \times 10^{22}$ cm$^{-2}$ (this does *not* include the partially ionized gas behind the hydrogen ionization front). This is much thicker than the classical Strömgren depth for planetary nebulae, because high energy photons can penetrate deeper than the classical depth and, at the high densities of BLRs, are able to ionize hydrogen from levels higher than the ground state. A general scaling derived from *Cloudy* models of BLR clouds is that $N_{ionized} \sim U \times 10^{23.7}$ cm$^{-2}$. The total mass is $M_{BLR} = 4\pi r^2 \, \mu_H \, m_H \, N_{ionized} \, \Omega/4\pi$, where $\Omega/4\pi$ is the covering factor required to reproduce the observed Lyα equivalent width, typically ~0.1. Since $4\pi r^2 = Q(H)/\Phi$, and using our assumption that $\lambda L_\lambda$ is constant between 1215 and 1450Å, we have as our new estimate $M_{BLR} \sim (6.67 \times 10^{13} \, L_{1450}/\Phi) \, (1450/1215) \, \mu \, m_H \, (10^{23.7} \, \Phi/n_H c) \, (\Omega/4\pi) = 160 \, (10^{11}/n_e) \, (L_{1450}/10^{44}) \, (\Omega/(0.4\pi))$ M$_{sun}$.

## 4. A More Realistic Estimate

The preceding mass estimates assume that the BLR gas is fully ionized, and are parameterized for a fiducial density ($10^{11}$ cm$^{-3}$) at which the gas emits with high efficiency. However, the reverberation results, when combined with the basic gas physics, show that neither of these things can be true. Figure 2 shows the Lyα emissivity of photoionized clouds per unit mass and per ionizing photon, as a function of the ionizing photon flux $\Phi$ and the gas density $n_H$. The fiducial density used above, $n_e \sim 10^{11}$ cm$^{-3}$, follows from combining the observed H$\beta$ luminosity from the Seyfert galaxy Arakelian 120 with the BLR size measured from the reverberation lag (Peterson et al. 1985). But there must also be gas with lower density, $n_e \sim 10^{9-10}$ cm$^{-3}$, in order to produce intercombination lines such as C III]. Therefore, there *must* be significant amounts of gas that emit with relatively low efficiency in lines like Ly$\alpha$, O VI, N V or C IV. In addition, the reverberation measurements show that the BLR is stratified into multiple emitting zones spread over a wide radial extent (with $R_{NV}/R_{MgII} \sim$ 20-30), corresponding to a range of roughly 3 decades in $\Phi$.

To find the total mass in a BLR that has a distribution in $\Phi$ and $n_H$, it is necessary to add up the contributions from all of the clouds at different positions on the log($\Phi$)–log($n_H$) plane. We do this for the *ionized* part of each cloud using the last method from the preceding section, summing up over all radii and densities the masses of individual shells $M_{shell} = 4\pi r^2 \mu_H m_H N_{ionized} (\Omega/4\pi)$. For $N_{ionized}$, we took the minimum of the stopping column density $N_{stop} = 10^{23}$ cm$^{-2}$ used in our *Cloudy* models, and $10^{23.7}U$, in order to exclude partially ionized gas lying beyond the ionization fronts in the individual clouds. This integration tells us the mass for the case $\Omega/4\pi = 1$. To find the covering factor and hence the actual mass required, we add up the Ly$\alpha$ equivalent width contribution from each shell and divide that into the observed Ly$\alpha$ equivalent width of 56Å.

Any viable model of the BLR must include gas dispersed over a wide region on the log($\Phi$)-log($n_H$) plane in order to reproduce both the reverberation results and the observed emission-line equivalent widths and intensity ratios. One approach to trying to identify the minimum $M_{BLR}$ consistent with the observed C III]/C IV intensity ratio is to use a single gas density $n_H = 10^{10}$ cm$^{-3}$, which is about the highest density at which C III] is still strongly emitted, and integrate over the necessary range in $\Phi$ indicated by the reverberation results. An example is Model F of Goad, O'Brien & Gondahlekar (1993), which requires $M_{BLR} = 700$ M$_{sun}$ when integrated using the technique described above.

However, the constant density model is unlikely to be correct. Rees, Netzer & Ferland (1989) pointed out that in such a model, if the BLR gas is in virialized motion over a large radial extent, the high and low-ionization emission lines would have very different profiles, contrary to the observations. Moreover, we have examples of individual Seyfert galaxies for which reverberation results show that the permitted lines are formed at densities $n_H \sim 10^{11}$ cm$^{-3}$, but C III] emission is seen that must be formed in regions of much lower density (e.g. NGC 5548; Ferland et al.1992).

This implies that within a BLR there is a wide range in $n_H$ as well as in the incident ionizing flux $\Phi$. The Locally Optimally-emitting Cloud (LOC) models (Baldwin et al. 1995) simulate this situation by assuming that the BLR is made up of a sea of emitting clouds spread out over the log($\Phi$)–log($n_H$) plane. Clouds naturally emit where $U$ takes on the correct value to produce a particular line, and it was shown that integrating over ensembles of clouds with power-law dependences of the numbers of clouds at different radii and different densities leads to spectra with line strengths and reverberation behavior (Korista & Goad, 2000) like those observed in AGN. Although the density distribution must be roughly proportional to $n_H^{-1}$ in order to produce the observed spectrum, the emission-line intensity ratios from the LOC models are remarkably insensitive to the exact radial distribution of the gas (Baldwin 1997).

In the standard LOC model from Baldwin et al. (1995), the radial and density distributions are chosen so that the number of clouds scales directly with both radius and density. The integration limits are $\log n_H \geq 8$, set by the absence of broad forbidden lines, and $\log(\Phi) \geq 18$, beyond which dust is no longer sublimated by the ionizing radiation. We also impose here an upper limit on $U$ in order to avoid counting in mass from clouds that are so highly ionized that they produce no UV or optical emission lines. We set this limit to be where EW(O VI $\lambda$1034) has dropped to 10% of its peak value. These limits are marked on Figure 2. Integrating over the $\log(\Phi)$-$\log(n_H)$ plane for clouds with stopping column densities $N_{stop} = 10^{23}$ cm$^{-2}$, we find that $M_{BLR} \sim 3600$ ($L_{1450}/10^{44}$) M$_{sun}$. If we raise the lower density limit of the LOC integration to $\log n_H \geq 9$, the BLR mass drops to $\sim$1000 M$_{sun}$, while if we integrate over the full $\log(\Phi)$-$\log(n_H)$ plane shown in Fig. 2 but below the limit in $U$ we get 20,000 M$_{sun}$. The minimum mass for this type of model is obtained by lowering the column density so that more of the clouds in the upper-left half of Figure 2 become optically thin in the Lyman continuum; $N_{stop} = 10^{22}$ cm$^{-2}$ is about the minimum column density that can still produce lines with ionization as high as O VI, and requires only 500 M$_{sun}$ if the minimum density is raised to $10^9$ cm$^{-3}$.

This confusing array of estimated $M_{BLR}$ are listed in Table 1. The problem is that the mass estimate depends strongly on how the gas is distributed over the $\log(\Phi)$–$\log(n_H)$ plane, but the computed intensity ratios between the UV emission lines (O VI, Ly$\alpha$, N V, C IV, He II, C III] and Mg II), which might be hoped to constrain the result, stay about constant. In fact, a strong argument in favor of the LOC model is that it explains why most QSO spectra are so similar to each other; it is because the observed spectrum is insensitive to the exact distribution of material in the BLR (see Baldwin 1997). Unfortunately, since there are many possibilities for the true gas distribution, the BLR mass must remain very uncertain. However, in all of the cases we have looked at, the mass must be several hundred to many thousands of solar masses for the high-redshift QSOs that we care about here. This is because for any viable model the locus of points on Figure 2 must include a great deal of gas that emits with relatively low efficiency per unit mass.

The lower limit on the mass that produces the observed emission lines is $M_{BLR} \sim 500$ ($L_{1450}/10^{44}$) M$_{sun}$, but this is for situations tuned to produce lines with maximum efficiency while still being compatible with the reverberation results. The models with column densities $N_{stop} = 10^{22}$ cm$^{-2}$ have been adjusted to just barely produce enough O VI $\lambda$1034, but are not able to produce the strong Ne VIII $\lambda$850 line seen in most QSO spectra that reach that far into the UV (see Hamann et al. 1998). The model with constant density $n_H = 10^{10}$ cm$^{-3}$ runs into the other difficulties mentioned above. A better estimate, for the less finely tuned situation that we suspect exists in nature, is $M_{BLR} \sim 10^3$–$10^4$ ($L_{1450}/10^{44}$) M$_{sun}$. The most luminous QSOs have $L_{1450} \sim 10^{45}$ erg s$^{-1}$ Å$^{-1}$, and thus $M_{BLR} \sim 10^4 - 10^5$ M$_{sun}$ for the ionized gas.

## 5. Discussion

Although the estimated BLR masses given in AGN textbooks range down to $M_{BLR} \sim 2$ M$_{sun}$, these are based on collisionally excited lines that come from only a small fraction of the BLR, on clouds at just one single density, on idealized line-emission coefficients that do not include photon destruction in actual clouds, and/or on luminosities characteristic of Seyfert galaxies rather than of high-redshift QSOs.

We find that when we take into account the large quantities of inefficiently-radiating gas that the reverberation results show must be present, the mass of the BLR in typical QSOs is at least 500 ($L_{1450}/10^{44}$) $M_{sun}$, and most likely several thousand solar masses with $M_{BLR} \sim 10^3$–$10^4$ ($L_{1450}/10^{44}$) $M_{sun}$. This estimate is for a BLR model that, unlike the single-zone models in the textbooks, is simultaneously consistent with the observed line ratios, equivalent widths, and reverberation behavior.

However, this is still just the line-emitting gas. It does not include the partially ionized or neutral gas behind the ionization fronts of the individual clouds. Nor does it include any gas in the large area covering the upper left corner of the log($\Phi$)-log($n_H$) plane, which does not emit any optical or ultraviolet emission lines and, if present, would correspond to the hot phase gas postulated by Krolik, McKee & Tarter (1981). Finally, there could easily be a large reservoir of neutral gas outside the BLR, shielded by dust beyond the radius for dust sublimation (see Netzer & Laor 1993) or in the dense accretion disk, again greatly increasing the mass of material that is dynamically related to or provides the source of material for the BLR. Thus, our estimate of $M_{BLR} \sim 10^3$-$10^4$ $M_{sun}$ is likely to still be just the tip of the iceberg in a situation where most of the gas connected with the BLR is invisible at optical – UV wavelengths.

The main question driving this paper is whether or not the BLR gas can tell us about the state of chemical evolution of the inner part of the host galaxy. Our results, combined with previous work on elemental abundances (e.g., Hamann & Ferland 1999), indicate that the BLR in luminous QSOs contains *at least* $10^3$–$10^4$ $M_{sun}$ of gas with a metallicity of solar or greater. Mass ejection from a single star or a few stars cannot cause this enrichment. A much larger stellar population in QSO host galaxies must be involved. We can place a firm lower limit on the mass of the stellar population contributing metals to the BLR by noting that this chemical enrichment is accompanied by locking gas up in stars and stellar remnants. As the system evolves, the fraction of the total mass that remains in gas declines as the metallicity increases. The exact relationship between the gas mass fraction and metallicity depends on the initial mass function (IMF), but for a broadly representative IMF (such as Salpeter 1955 or Scalo 1986) the gas fraction is only about 10% to 20% when the gas-phase metallicity reaches solar. (We could draw the same conclusion by noting simply that in our own Galaxy the gas fraction in the solar neighborhood is of order 10% while the gas-phase metallicity is roughly solar.) Therefore, $\geq 10^3$–$10^4$ $M_{sun}$ of BLR gas with solar or higher metallicity requires enrichment by a stellar population with mass $\geq 10^4$–$10^5$ $M_{sun}$.

However, this mass limit is probably a gross underestimate of the actual stellar populations probed by BLR metal abundances. In particular, we assumed above that the entire interstellar medium enriched by the stars appears in the BLR. Any realistic scenario for the fuelling of QSOs and the migration of gas toward galactic nuclei involves far more gas than appears in the BLR. The mass of the accretion disk around a $10^8$ $M_{sun}$ black hole is likely to be $\sim 10^6$ – $10^7$ $M_{sun}$.[1] Molecular tori surrounding AGN nuclei can be another significant gas component, at least in low-luminosity AGN (for example, the pc-scale water megamasers seen in several Seyfert galaxies contain about $10^3$–$10^5$ $M_{sun}$ of molecular gas; Maloney 2002, Gallimore et al. 1996), although it is not clear how this mass might scale in the presence of a luminous QSO). However, the major repository of inflowing material must be the central black hole itself, which for the fiducial QSO discussed here should have a mass of order $10^8$-$10^9$ $M_{sun}$ (e.g., Kaspi et al. 2000; Peterson & Wandel 2000; Ferrarese et al. 2001; Vestergaard 2002) – far greater than the estimated BLR mass.

---

[1] The former estimate is for a thick disk (Madau 1988). The latter estimate is for a thin disk, following Frank, King & Raine (1992, p. 199), integrating out to 1000 $R_{gravitational}$ with the viscosity parameter $\alpha_{visc} = 0.1$, and calculating $dM/dt$ from $L_{1450} = 10^{44}$ erg s$^{-1}$ Å$^{-1}$ using the above spectral energy distribution and a mass-to-energy conversion efficiency $\eta = 0.1$.

Therefore, considerably more mass is funneled into the central regions than that which appears in the BLR at any given time. In addition, star clusters with masses of order $10^5$ or even $10^6$ $M_{sun}$, reminiscent of globular clusters, do not reach solar or higher metallicities in isolation. A much deeper gravitational potential is needed to retain the gas against the building thermal pressures caused by supernovae. The massive host galaxies of QSOs provide this gravitational potential. In these environments, it is unrealistic to suppose that the prompt enrichment to high metallicities occurs in only a small central star cluster with mass of order $\sim 10^5$ $M_{sun}$. This environment should be overwhelmingly diluted by the influx of metal poor gas from the surrounding galaxy. Specific models of the chemical evolution predict, instead, that the rapid rise to BLR-like metallicities occurs on much larger scales in galactic nuclei, involving bulge-size stellar populations with masses $\geq 10^9$ $M_{sun}$ (Friaca & Terlevich 1998).

We conclude that the BLR does sample gas from a significant portion of the inner regions of the host galaxy. Quasar emission lines can thus reveal the chemical evolution and physical state of gas in the very center of the most massive galaxies. The tools needed to exploit them are now available. The line intensity ratios already indicate that the chemical composition of the emitting gas correlates with quasar luminosity in a way that is suggestive of the known galactic mass/luminosity/metallicity correlations (Hamann & Ferland 1993, 1999). The metallicities are at or above solar even at the highest redshifts, showing that the quasar phenomenon must not begin until ~0.1 Gyr after stellar evolution starts. We have shown here that these results derived from the broad emission lines have general significance for the evolution of galaxies.

We gratefully acknowledge NASA grant HST-GO-08283 for support of this research project. FH acknowledges financial support from the National Science Foundation through grant AST 99-84040. GJF thanks the NSF (AST 0071180) and NASA (NAG5-8212 and NAG5-12020) for support.

# References

Baldwin, J., Ferland, G., Korista, K. & Verner, D. 1995, ApJL, 455, L119

Baldwin, J.A. 1997, in IAU Colloq. 159, Emission Lines in Active Galaxies: New Methods and Techniques, ed. B.M.Peterson, F.-Z. Cheng & A.S.Wilson (ASP Conf. Ser. 113; San Francisco:ASP), 80

Boyle, B. J., Shanks, T., Croom, S. M., Smith, R. J., Miller, L., Loaring, N. & Heymans, C. 2000, MNRAS, 317, 1014.

Dietrich, M., Hamann, F., Shields, J.C., Constantin, A., Junkkarinen, V.T., Chaffee, F. & Foltz, C.B. 2002, ApJ (in press)

Ferland, G.J., Peterson, B.M., Horne, K., Welsh, W.F. & Nahar, S.N. 1992, ApJ, 387, 95

Ferland, G.J. 2002 Hazy, a brief introduction to Cloudy 96.00, (Univ of Kentucky Internal Report)

Ferrarese, L., Pogge, R.W., Peterson, B.M., Merritt, D., Wandel, A., & Joseph, C.L. 2001, ApJ, 555, L79

Friaca, A.C.S. & Terlevich, R.J. 1998, MNRAS, 298, 399

Frank, J., King, A.& Raine, D. 1992, Accretion Power in Astrophysics, 2nd Edition (Cambridge, U.K.: Cambridge University Press)

Gallimore, J.F., Baum, S.A., O'Dea, C.P., Brinks, E. & Pedlar, A. 1996, ApJ, 462, 740

Goad, M.R., O'Brien, P.T. & Gondhalekar, P.M. 1993, MNRAS, 263, 149

Hamann, F & Ferland, G. 1993, ApJ, 418, 11

Hamann, F & Ferland, G. 1999, Ann.Rev.A&A, 37, 487

Hamann, F.,Cohen, R.D., Shields, J.C., Burbidge, E. M., Junkkarinen, V. & Crenshaw, D. M. 1998, ApJ, 496, 761

Hewett, P.C., Foltz, C.B. & Chaffee, F.H. 1995, AJ, 109, 1498

Kaspi, S., Smith, P.S., Netzer, H., Maoz, D., Jannuzi, B.T., & Giveon, U. 2000, ApJ, 533, 631

Kauffmann, G.& Haehnelt, M. 2000, MNRAS, 311, 576

Korista, K., Baldwin, J., Ferland, G. and Verner, D. 1997, 108, 401

Korista, K.T. & Goad, M.R. 2000, ApJ, 536, 284

Krolik, J.H., McKee, C.F. & Tarter, C.B. 1981, ApJ, 249, 422

Krolik, J.H. 1999, Active galactic nuclei : from the central black hole to the galactic environment ( Princeton, N. J. : Princeton University Press)

Kwan, J. & Krolik, J.H. 1981, ApJ, 250, 478

Madau, P. 1988, ApJ, 327, 116

Maloney, P. 2002, astro-ph/0203496

Netzer, H. & Laor, A. 1993, ApJ, 404, L51

Osterbrock, D.E. 1989, Astrophysics of Gaseous Nebulae and Active Galactic Nuclei (Sausalito: University Science Books)

Peterson, B.M., Meyers, K.A., Capriotti, E.R., Foltz, C.B., Wilkes, B.J. & Miller, H.R. 1985, ApJ, 292, 164

Peterson, B.M. 1993, PASP, 105, 247

Peterson, B.M. 1997, An Introduction to Active Galactic Nuclei (Cambridge: Cambridge University Press).

Peterson, B.M. & Wandel, A. 2000, ApJ, 540, L13

Rees, M.J., Netzer, H. & Ferland, G.J. 1989, ApJ, 347, 640.

Salpeter, E.E. 1955, ApJ, 121, 161

Scalo, J.M. 1986, Fund. Cos. Phys., 11, 1

Silk, J. & Rees, M.J. 1998, A&A, 331, 1

Vestergaard, M. 2002, ApJ, 571, 733

| Table 1. BLR Mass Estimates for Various Models, for QSOs with $L_{1450} = 10^{44}$ erg s$^{-1}$ Å$^{-1}$ | |
|---|---|
| **Model** | $M_{BLR}$ ($M_{sun}$) |
| **Single-Cloud Models with $n_H = 10^{11}$ cm$^{-3}$ (§3)** | |
| From Lyα emissivity for case B | 29 |
| From Lyα emissivity including Lyα thermalization & destruction | 100 |
| From mass within the ionized depth | 160 |
| **LOC models (§4)** | |
| $N_{stop} = 10^{22}$ cm$^{-2}$, $\log(n_H) \geq 9$, $\log(\Phi) \geq 18$ | 500 |
| $N_{stop} = 10^{23}$ cm$^{-2}$, Constant density $\log(n_H) = 10$ (Goad et al. Model F) | 700 |
| $N_{stop} = 10^{23}$ cm$^{-2}$, $\log(n_H) \geq 9$, $\log(\Phi) \geq 18$ | 1000 |
| $N_{stop} = 10^{23}$ cm$^{-2}$, $\log(n_H) \geq 8$, $\log(\Phi) \geq 18$ (standard LOC model) | 4000 |
| $N_{stop} = 10^{23}$ cm$^{-2}$, $\log(n_H) \geq 7$, $\log(\Phi) \geq 17$ | 20000 |

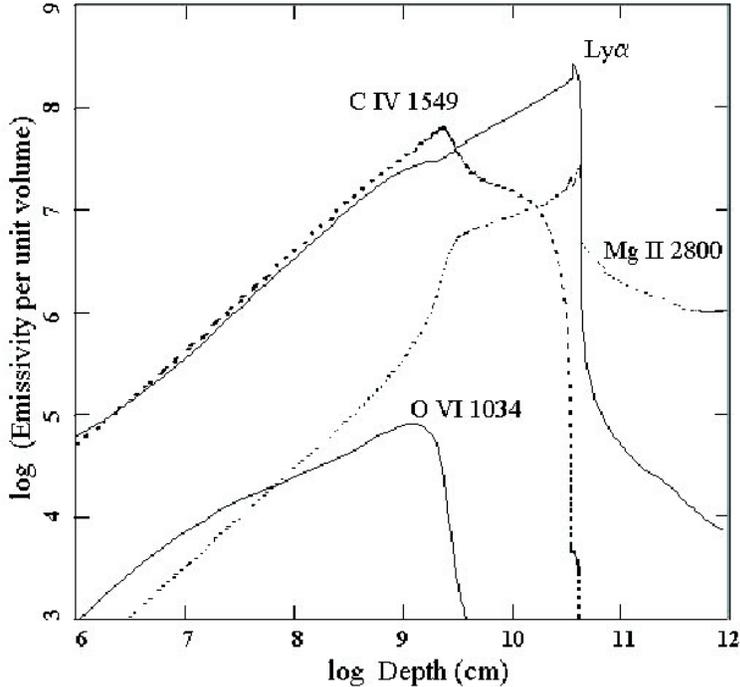

**Figure 1.** The emergent intensity per unit volume of several emission lines (erg cm$^{-3}$ s$^{-1}$), vs. the depth $D$ into our fiducial BLR cloud. Note the logarithmic scales. The CIV line emissivity peaks at $D \sim 2\times10^9$ cm, while Lyα emits strongly over a 16 times greater range in depth.

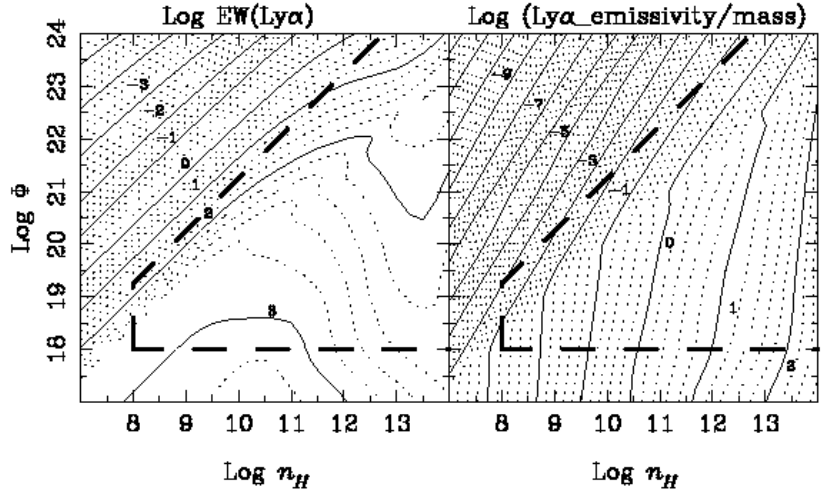

**Figure 2.** (left panel) Logarithmically spaced contours of Lyα equivalent width as a function of ionizing photon flux $\Phi$ and gas density $n_H$. This shows the rapid drop in the efficiency with which continuum photons are converted to Lyα for $n_H > 10^{11}$ cm$^{-3}$.

(right panel) Logarithmically spaced contours of Lyα emission efficiency per unit mass, normalized to the value for our fiducial model at $\log(\Phi) = 19.5$, $\log(n_H) = 11$. To find the minimum BLR mass if all of the BLR gas has some given value of $\Phi$ and $n_H$, the mass $M_{BLR} = 160\,(L_{1450}/10^{44})$ M$_{sun}$, corresponding to our fiducial model, should be divided by the factor given on this figure. The heavy dashed lines enclose the area included in the standard LOC model, which equally weights clouds at all positions on the $\log(\Phi)$-$\log(n_H)$ plane. The reverberation observations directly show that the gas is distributed over a factor ~30 in radius, or 3 decades in $\log(\Phi)$, and an appreciable amount must have $n_H \leq 10^{10}$ cm$^{-3}$ to produce C III] . For any compatible model, much of the gas must emit with low Lyα efficiency, and therefore a great deal of extra mass is required.